\documentclass[showpacs,aps,floatfix,prx,reprint,superscriptaddress]{revtex4-1} 

\usepackage{xfrac}
\usepackage{amssymb}
\usepackage{braket}
\usepackage{graphicx}
\usepackage{verbatim}
\usepackage{etoolbox}
\usepackage{amsfonts} 
\usepackage{amsmath} 
\usepackage[utf8]{inputenc} 
\usepackage{mathtools}
\usepackage{soul,xcolor,colortbl}
\usepackage{setspace}
\usepackage{multirow}
\usepackage{hyperref} \hypersetup{colorlinks=true,citecolor=blue,linkcolor=blue,urlcolor=blue} 
\usepackage{makecell}
\usepackage{ifthen}

\usepackage{bm} 
\usepackage{array}
\newcolumntype{C}[1]{>{\centering\arraybackslash}p{#1}}\usepackage{soul}
\definecolor{Gray}{gray}{0.85}

\usepackage{color} 
\usepackage{morefloats} 

\definecolor{Gray}{gray}{0.9}
\definecolor{LightCyan}{rgb}{0.88,1,1}

\def\AFLOW{{\small AFLOW}}
\def\AFLUX{{\small AFLUX}}
\def\LUX{{\small LUX}}

\def\AEL{{\small AEL}}
\def\AGL{{\small AGL}}
\def\API{{\small API}}
\def\URI{{\small URI}}

\def\REST{{\small REST}}

\def\GUI{{\small GUI}}
\def\SQL{{\small SQL}}
\def\ABNF{{\small ABNF}}
\def\AUID{{\small AUID}}
\def\AURL{{\small AURL}}

\def\citeAFLOWAGL{\cite{curtarolo:art96, curtarolo:art115}}
\def\citeAFLOW{\cite{aflowPAPER,curtarolo:art110,curtarolo:art104, curtarolo:art63,curtarolo:art57,curtarolo:art49,monsterPGM}}
\def\citeAFLOWML{\cite{curtarolo:art84,curtarolo:art85,curtarolo:art94,curtarolo:art120}}
\def\citeAFLOWLIB{\cite{aflowlibPAPER,curtarolo:art58,curtarolo:art92}}

\begin{document} 



\title{\LARGE {AFLUX: The LUX materials search API \\ for the AFLOW data repositories}}

\author{Frisco Rose}  \affiliation{Department of Mechanical Engineering and Materials Science, Duke University, Durham, North Carolina 27708, USA}
\author{Cormac Toher} \affiliation{Department of Mechanical Engineering and Materials Science, Duke University, Durham, North Carolina 27708, USA}
\author{Eric Gossett}    \affiliation{Department of Mechanical Engineering and Materials Science, Duke University, Durham, North Carolina 27708, USA}
\author{Corey Oses}     \affiliation{Department of Mechanical Engineering and Materials Science, Duke University, Durham, North Carolina 27708, USA}
\author{Marco Buongiorno Nardelli}  \affiliation{Department of Physics and Department of Chemistry, University of North Texas, Denton TX}
\author{Marco Fornari}                      \affiliation{Department of Physics, Central Michigan University, Mount Pleasant, MI 48858, USA }
\author{Stefano Curtarolo}   \email[]{stefano@duke.edu} \affiliation{Department of Mechanical Engineering and Materials Science, Duke University, Durham, North Carolina 27708, USA}

\begin{abstract}
  Automated computational materials science frameworks rapidly generate large quantities of materials data useful for accelerated materials design.
  We have extended the data oriented \AFLOW-repository \API\ (\underline{A}pplication-\underline{P}rogram-\underline{I}nterface,
  as described in Comput. Mater. Sci. {\bf 93}, 178 (2014)) to enable programmatic access to search queries.
  A \URI-based search \API\ (\underline{U}niform \underline{R}esource \underline{I}dentifier) is proposed for the construction of complex queries
  with the intent of allowing the remote creation and retrieval of customized data sets.
  It is expected that the new language \AFLUX, acronym for \underline{A}utomatic \underline{F}low of \underline{LUX} (light),
  will facilitate the creation of remote search operations on the \AFLOW.org set of computational materials science data repositories.
\end{abstract}

\date{\today} 
\maketitle 


\section{Introduction}
\label{introduction}




Automated computational materials science frameworks such as \AFLOW\ \citeAFLOW\ rapidly 
generate large quantities of materials data, which can be used to identify trends, build machine 
learning models \citeAFLOWML\ and finally accelerate materials design \cite{curtarolo:art81}.
The dissemination of such large quantities of data requires the use of online advanced databases such as
\AFLOW.org \citeAFLOWLIB, NoMaD \cite{nomad}, Materials Project \cite{materialsproject.org}, 
{\small OQMD} \cite{Saal_JOM_2013} and AiiDA \cite{aiida.net, Pizzi_AiiDA_2016} to make
the data accessible in an effective fashion. 

The \AFLOW\ data repositories \citeAFLOWLIB\ currently contains in excess of 1.5 million materials entries, and over
150 million materials properties (keywords) which are directly accessible using the existing \AFLOW\ data
\REST-\API\ \cite{curtarolo:art92}.  
The existing data \API\ presents data effectively, but it does not enable programmatic access to the types of complex search queries
available through the \AFLOW\ online web portal. 
In order to address these issues, we have implemented a \URI-based search \API\ (\underline{U}niform \underline{R}esource \underline{I}dentifier)
which enables the construction of complex search queries and
allows for the creation and retrieval of customized data sets.

For large, automatically generated datasets such as those contained in the \AFLOW\ repository, 
automated verification procedures for calculation quality becomes an important issue, as it is 
impossible for a user to individually check the calculations for every single entry in a data set.
The \AFLUX\ search \API\ can be used to help facilitate such verification checks, since it is possible
to include properties which provide information about calculation
quality in the results returned by a search query, or even
to filter the results to remove entries which do not fulfill specified convergence criteria. To facilitate
this, the \AFLOW\ data \REST-\API\ has been extended to include additional keywords to
expose information concerning the calculation settings and convergence. This upgraded version 
of the data \API\ is termed version 1.2 as compared to version 1.0 described in Ref. \onlinecite{curtarolo:art92}.
Even and odd version numbers correspond to stable and developmental iterations, respectively.

\section{\AFLUX\ Search \API}
\label{AFLOWLux}

Access to relational information is typically a highly involved process that
requires specialized knowledge not generally available or readily attainable by
general consumers of said information. An access mechanism that exposes the
data to human directed exploration via a \underline{g}raphical \underline{u}ser \underline{i}nterface (\GUI) is
typically limited by the \GUI\ designers imagined end use of the data, and is
practically impossible to access via programmatic methods.  Exposing data via a
procedural approach, such as using \underline{S}tructured \underline{Q}uery \underline{L}anguage (\SQL), requires
specialized knowledge of both the language in use and the (often highly
convoluted) organization of the underlying data. In order to address these
concerns we created a domain agnostic text based search \underline{a}pplication
\underline{p}rogramming \underline{i}nterface (\API) named \LUX.  We present
\LUX\ in the \AFLOW\ domain context as \AFLUX\, with the design goals of
exposing an \API\ that is human accessible, logically robust, concise and as
lucid as possible. 

\subsection{Design Specification}

The design of this \API\ is driven by the need for a straightforward and globally
accessible representation of the \AFLOW\ search features. The feature set
should reflect and extend the capabilities found in the \AFLOW\ online search
\GUI\ available at \url{aflow.org}. Further motivation comes from the attempt to
unburden the end user from needing to understand the intricacies of \SQL\ or a
particular database schema.  An additional benefit of eliminating the explicit
representation of the database ({\small DB}) schema is that we are free to optimize the
internal {\small DB} structure without breaking backwards compatibility in the \AFLUX\
\API. Although \LUX\ is designed to operate in an origin agnostic fashion, we
must make concessions to older software stacks that may be unable to create an
extended length \URI, and thus \LUX\ strives to balance conciseness with human
readability. As a final consideration, the \LUX\ syntax should be generally
extensible, both to accommodate the growth of the underlying data and to allow
\LUX's application to an arbitrary data store. The \LUX\ syntax was conceived
out of these requirements, and for the purpose of this presentation \LUX\ will
be described in the \AFLOW\ context.  The \LUX\ \API\ language construct is
superficially like a C/C++ style subroutine / function call, which we call a
matchbook. The utility of \LUX\ is derived from the fact that the
intra-matchbook restrictions and inter-matchbook declarations are logically
related search criteria.

\subsection{\URI\ Query Context}

The implementation of the \LUX\ search \API\ will use an internet \URI\ as the
submission layer. In a related effort, the \AFLOW\ data \API\ also uses a \URI\ to
access large data stores \cite{curtarolo:art92}. However, in the data \API\ the path portion of the \URI\
is used to uniquely identify each material and navigate between materials,
while the query portion of the \URI\ is used to isolate specific data values and
dictate the associated data response format. In using the \LUX\ \API\ to summons
data the query portion of a \URI\ is more suited to the highly dynamic nature of
\LUX{}'s relational criteria.

The beginning of the query section in the \URI\ is
defined by the presence of the first question mark character ``?'' in the \URI.
The query section is terminated by either the presence of the fragment section,
indicated by the first octothorpe character ``\#'', or by the end of the \URI.

Thus the standards for the query portion of a \URI\ forms the foundation of the
\LUX\ \API.


\subsection{Summons Syntax}

\begin{figure*}[t!]
\includegraphics[width=0.99\textwidth]{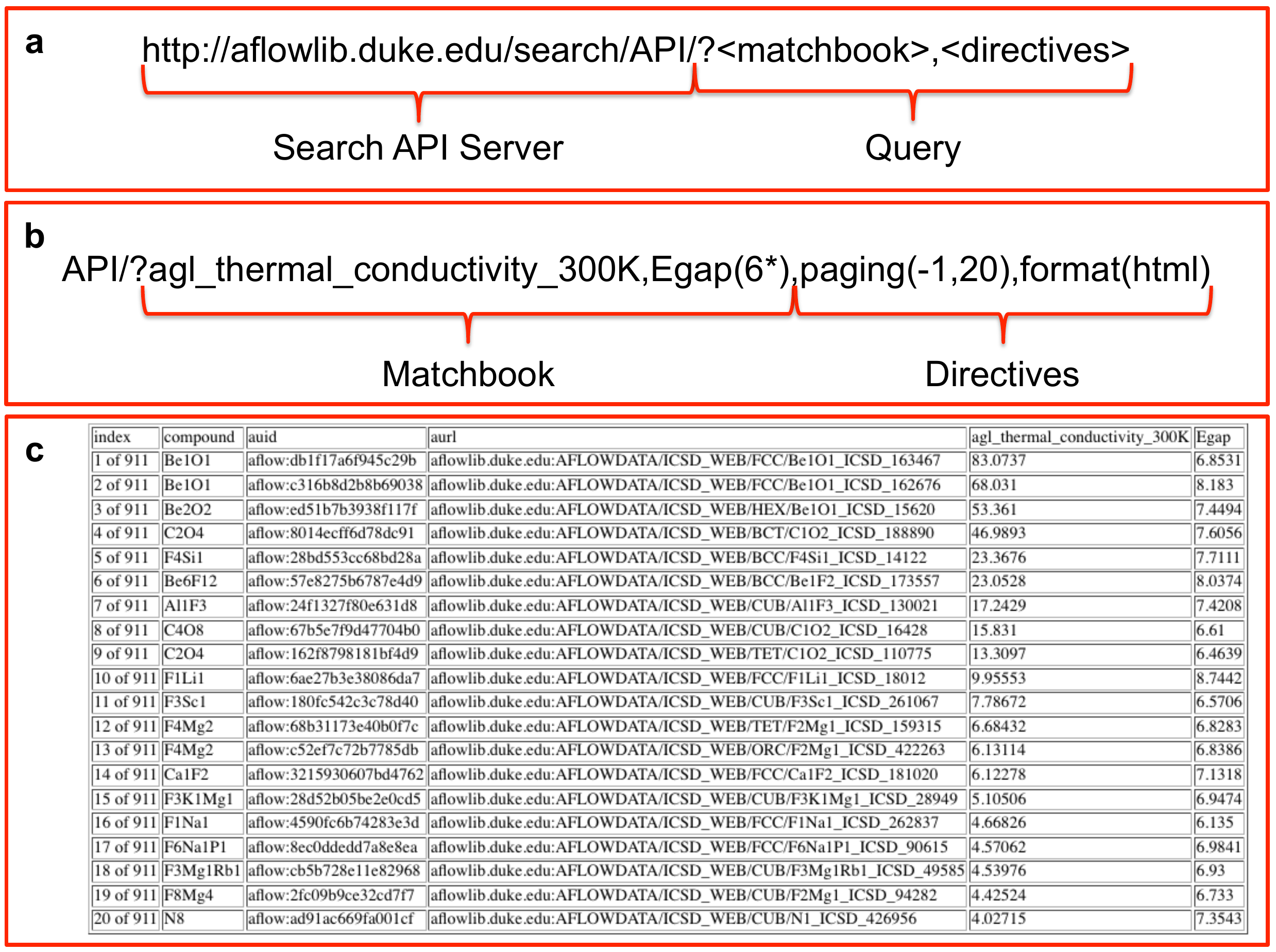}
\vspace{-2mm}
\caption{
\small 
{\bf (a)} \URI\ format for \LUX\ search query. Note that the entire Matchbook must be before
all of the Directives.
{\bf (b)} Example search query for a material to act as an electrically insulating heat sink.  
{\bf (c)} Results returned in HTML format.
}
\label{fig:search_format_example}
\end{figure*}

The summons syntax is broken into two segments with \URI\ query safe characters
forming the body of the summons, as shown in Fig.
\ref{fig:search_format_example}(a).  The first portion contains the relational
property match criterion with its respective matching restrictions - this is
the Matchbook. The Matchbook portion is followed by the Directive portion with
its attendant attributes. While there is no {\it a priori} order dependency
within the Matchbook or the Directives, the Matchbook must come entirely
before the Directives in the submitted \URI, as shown in Fig.
\ref{fig:search_format_example}.

The property matchbook portion is mandatory for a meaningful data response. In
contrast a directive only summons may return metadata or an error state,
depending on the directive of concern and the manner in which is is summoned.

\subsection{Character Set}

As one of the primary design objectives of \LUX\ is to be human readable, we
must consider the allowed \URI\ characters in that context. The \URI\ generic syntax
({\small RFC3986} \cite{RFC3986}) defines the explicit restrictions on \URI\ query
formation in \ABNF\ notation \cite{RFC5234}. As we want our search syntax to be
human readable, we will limit our overt use of percent (pct) encoded
characters. Finally, in order to support a concise summons we have opted to
forgo verbose operators and use single character operators. What follows is a
derivation of the \LUX\ character set based on our design goals and the
limitations of the standards that \LUX\ will operate under. 


From section 3.4 of Ref. \onlinecite{RFC3986} the query section of the \URI\
allows for the use of the following characters
\begin{quote}
query = *( pchar / ``/'' / ``?'' )
\end{quote}
which can be read as follows: allow zero or more of the characters from the
conjunction of the pchar set, the solidus character ``/'', and the question mark
character ``?'', where the ``pchar'' character set is defined in section 3.4 of 
Ref. \onlinecite{RFC3986} as
\begin{quote}
pchar = unreserved / pct-encoded / sub-delims / ``:'' / ``@''.
\end{quote}
``Unreserved'' is defined in section 2.3 of Ref. \onlinecite{RFC3986} as
\begin{quote}
unreserved = ALPHA / DIGIT / ``-'' / ``.'' / ``\_'' / ``\textasciitilde{}'',
\end{quote}
and ``sub-delims'' are defined in Section 2.2 of Ref. \onlinecite{RFC3986} as
\begin{quote}
sub-delims = ``!'' / ``\$'' / ``\&'' / `` ' '' / ``('' / ``)'' / ``*'' / ``+'' / ``,'' / ``;''
/ ``=''.
\end{quote}

So, in summary, a strict superset of the characters that our syntax may
be constructed from is as follows:
\begin{quote}
ALPHA / DIGIT / ``-'' / ``.'' / ``\_'' / ``\textasciitilde{}''/ ``!'' / ``\$'' / ``\&'' / ``''' / ``('' / ``)'' / ``*'' / ``+'' / ``,'' / ``;'' / ``='' /``:'' / ``@'' /``/'' / ``?''.
\end{quote}
However, while this is the latest relevant RFC there are older protocols
implemented by browsers/servers/languages that interfere with the query
character set. For instance, RFC1630 and RFC2396 reserved the plus sign.  We
will attempt to avoid any characters that may result in an unintended mangling
of the query.

\subsection{Operator Character Codex}

We are not explicitly limited by any origin standards, such as {\small HTML5} which only
allows unreserved and pct-encoded, as the origin of the query is application
agnostic and only requires the ability to construct a valid \URI. As an aside,
it is often possible to work around the limitation of origin standards by
calling lower level methods in the application stack. In essence we need only
open a bi-directional communication socket to the \LUX\ \API, send an un-mangled
\URI\ with an appropriate query structure to the server, and accept the returned
data set. That said, we want to make the implementation as easy, unambiguous
and unimpeded as we can. As a first consideration we will avoid the use of
``?'', ``\&'', ``+'', ``;'' and ``='' in order to avoid confusing arbitrarily
handled webserver/cgi/PHP/Web-Browser/... \URI\ parsers.  This will also help
distinguish \LUX\ from preconceived end user expectations of other, more
simplistic, query based \API\ behaviors. We also reserve ``.'' and ``-'' as they
are inherent to the definition of a number. Furthermore, we want to use ALPHA/DIGIT
as general string/number characters. Lastly we reserve ``\_'' as we use it in
the keyword (\AFLOW\ Snake Case keywords for instance) identifier amalgamation
convention.  This leaves

\begin{quote}
``\textasciitilde{}'' / ``!'' / ``\$'' / `` ' '' / ``('' / ``)'' / ``*'' / ``,'' / ``:'' /
``@'' / ``/''.
\end{quote}

Thus we define the \LUX\ operator tokens as shown in Table \ref{tab:lux_codex}.

\begin{table*}[t!]
  \caption{\small \LUX\ Codex. List of the symbols used to represent logic operations in the \LUX\ Search-\API\ syntax. 
Note that three symbols are currently reserved for future use.
  }
  \label{tab:lux_codex}
  {\footnotesize
    \begin{tabular}{| l | c | l |}
      \hline
     Logic operation  & Operator symbol & Description \\
      \hline
      BLOCK-NEW & ( & The new set precedence context \\ \hline
      BLOCK-END & ) & The end set precedence context \\ \hline
      UNARY-MUTE & \$ & The property output suppression operator \\ \hline
      UNARY-NOT & ! & The logical inversion unary operator \\ \hline
      UNARY-LOOSE & * & The positional loose datum match unary operator \\ \hline
      BINAL-AND & , & The logical conjunction binary operator/list separator \\ \hline
      BINAL-OR & : & The logical disjunction binary operator/list separator \\ \hline
      DATUM-STR & ' & The explicit string datum context \\ \hline
      RESERVED & @ & Reserved for future use \\ \hline
      RESERVED & \textasciitilde{} & Reserved for future use \\ \hline
      RESERVED & / & Reserved for future use \\
      \hline
    \end{tabular}
  }
\end{table*}

\begin{table*}[t!]
  \caption{\small Numerical filtering.
  }
  \label{tab:numerical_filter}
  {\footnotesize
    \begin{tabular}{| c | c |}
      \hline
      Query syntax  & Result description \\
      \hline
      \textless{}keyword\textgreater{}(1.234) & \makecell{Return only results where value in  \textless{}keyword\textgreater{} equals 1.234} \\ \hline
      \textless{}keyword\textgreater{}(!1.234) & \makecell{Return only results where value in  \textless{}keyword\textgreater{} does not equal 1.234} \\ \hline
      \textless{}keyword\textgreater{}(*1.234) & \makecell{Return only results where value in \textless{}keyword\textgreater{} is less than or equal to 1.234} \\ \hline
      \textless{}keyword\textgreater{}(!*1.234) & \makecell{Return only results where value in \textless{}keyword\textgreater{} is not less than or equal to 1.234, i.e. is greater than 1.234} \\ \hline
      \textless{}keyword\textgreater{}(1.234*) & \makecell{Return only results where value in  \textless{}keyword\textgreater{} is greater than or equal to 1.234} \\ \hline
      \textless{}keyword\textgreater{}(!1.234*) & \makecell{Return only results where value in \textless{}keyword\textgreater{} is not greater than or equal to 1.234, i.e. is less than 1.234} \\
      \hline
    \end{tabular}
  }
\end{table*}







\subsection{Properties are Reserved Words}

In addition to the character limitations as defined by the operator character
codex there exist reserved words. These words fall into two categories: domain
specific (e.g. \AFLOW\ properties) keywords and \LUX\ directives. The list of
the \AFLOW\ property data keywords valid as of this draft are those keywords in
the original \AFLOW\ data \REST\ \API\ \cite{curtarolo:art92} and recently
developed keywords given in Appendix \ref{table_properties}.

\subsection{Directives are Pseudo Property Keywords}

\begin{table*}[t]
  \caption{\small String filtering.
  }
  \label{tab:string_filter}
  {\footnotesize
    \begin{tabular}{| c | c |}
      \hline
      Query syntax  & Result description \\
      \hline
      \textless{}keyword\textgreater{}('foo') & \makecell{Return only results where the string in  \textless{}keyword\textgreater{} is exactly ``foo''} \\ \hline
      \textless{}keyword\textgreater{}(!'foo') & \makecell{Return only results where the string in  \textless{}keyword\textgreater{} is not ``foo''} \\ \hline
      \textless{}keyword\textgreater{}(*'foo') & \makecell{Return only results where the string in  \textless{}keyword\textgreater{} ends with ``foo''} \\ \hline
      \textless{}keyword\textgreater{}(!*'foo') & \makecell{Return only results where the string in  \textless{}keyword\textgreater{} does not end with ``foo''} \\ \hline
      \textless{}keyword\textgreater{}('foo'*) & \makecell{Return only results where the string in  \textless{}keyword\textgreater{} starts with ``foo''} \\ \hline
      \textless{}keyword\textgreater{}(!'foo'*) & \makecell{Return only results where the string in  \textless{}keyword\textgreater{} does not start with ``foo''} \\ \hline
      \textless{}keyword\textgreater{}(*'foo'*) & \makecell{Return only results where the string in  \textless{}keyword\textgreater{} contains the substring ``foo''} \\ \hline
      \textless{}keyword\textgreater{}(!*'foo'*) & \makecell{Return only results where the  string in \textless{}keyword\textgreater{} does not  contain the substring ``foo''} \\
      \hline
    \end{tabular}
  }
\end{table*}

\begin{table*}[t!]
  \caption{\small Multiple criteria filtering.
  }
  \label{tab:multiple_filter}
  {\footnotesize
    \begin{tabular}{| l | l |}
      \hline
      Query syntax  & Result description \\
      \hline
      \textless{}keyword\textgreater{}(1*,*1.234) & \makecell{Returns only results where value in \textless{}keyword\textgreater{} is between 1.0 and 1.234} \\ \hline
      \textless{}keyword\textgreater{}(1*),\textless{}keyword\textgreater{}(*1.234)  & \makecell{Logically equivalent to \textless{}keyword\textgreater{}(1*,*1.234)} \\ \hline
      \textless{}keyword\_a\textgreater{}(1.234), \textless{}keyword\_b\textgreater{}(`foo') & \makecell{Return results that match for \textless{}keyword\_a\textgreater{}(1.234) and \textless{}keyword\_b\textgreater{}(`foo')} \\ \hline
      \textless{}keyword\_a\textgreater{}(1.234) : \textless{}keyword\_b\textgreater{}(`foo') & \makecell{Return results that match for \textless{}keyword\_a\textgreater{}(1.234) or \textless{}keyword\_b\textgreater{}(`foo')} \\
      \hline
    \end{tabular}
  }
\end{table*}

\LUX{}'s directives do not necessarily have an explicit presence in the output
data. Their main purpose is to provide context and affect the way that the
output is produced.

The list of directives is intentionally short and is as follows:
\begin{quote}
{[}``catalog'', ``format'' ,``help'' , ``paging'', ``schema''{]}.
\end{quote}

{\bf \hfill Directive: catalog \hfill}

It is often helpful to provide a higher-level abstraction of the data sets
available. In \AFLOW\ we need to control which library or libraries of
materials we wish to query. The internal implementation of the catalog
directive could have been provided as a domain specific keyword if the
defining property was guaranteed to be ubiquitous, however this may not be
possible for all domains and thus ``catalog'' has been designated in the
directive context to allow for specialized internal handling. Within \AFLOW\,
the returned set of material properties can be restricted by the use of the
``catalog'' directive.  The currently available catalogs are:
{[}``icsd'',``lib1'',``lib2'',``lib3''{]}.

By default all catalogs are included in a search request and hence the catalog
directive may be omitted in many cases. The principle use scenario for catalog
is to restrict the search to a subset of endpoint resources. The catalog
directive is called as: 

\url{API-Server?<Matchbook>,catalog(library\_a)}

for just ``library\_a''; or as 

\url{API-Server?<Matchbook>,catalog(library\_a:library\_b)}

to search inclusively in both libraries ``a'' and ``b'' (i.e. an entry can be
in the conjunction of ``a'' and ``b'').

{\bf \hfill Directive: format \hfill}

We need to be able to format output of the resultant data. The currently
supported output modifiers are ``html'' and ``json''.  Use the ``format''
keyword to set the output type. The default response is compact JSON, setting
``json'' explicitly returns a more legible JSON response.

\url{API-Server?<Matchbook>,format(json)}

or

\url{API-Server?<Matchbook>,format(html)}

{\bf \hfill Directive: help \hfill}

The help directive with no arguments will return a response with a summary of
the proper syntax and use of \LUX{}. If a specific help context has been
internally defined for a keyword, then the keyword may be passed to the help
directive to retrieve the associated message, otherwise the response will be
the same as if no keyword had been supplied.

\url{API-Server?help()}

or

\url{API-Server?help(keyword)}

{\bf \hfill Directive: paging \hfill}

We need the ability to control the number of responses as some searches may
return exceedingly large data sets (potentially exceeding tens of millions of
data points). By default we will assume 64 data sets in a response page. This
return limit can be overridden with the ``paging'' directive. In the following
examples the $n^{\mathrm{th}}$ page of $k$ data sets can be requested. Paging
has special powers; asking for zero results ($k=0$) will return the number of
matching data sets; asking for the zeroth page ($n=0$) will return all matching
data sets regardless of $k$ - \emph{\textbf{be careful}} as this can be a large
response! Negating $n$ will change the sort order to descending. The paging
directive is called as:

\url{API-Server?<Matchbook>,paging(n)}

or

\url{API-Server?<Matchbook>,paging(n,k)}

{\bf \hfill Directive: schema \hfill}

The schema directive allows external access to the internal data store
metadata. The returned metadata represents useful information about the
internal nature of the keyword data allowing an end user to make informed
decisions about how to handle the property or properties of interest. The \LUX\
language makes no assumptions regarding the metadata content. This allows
freedom in the way that \LUX\ is implemented on the backing store. For the \AFLUX\
instance we expose the \AFLOW\ Data schema to allow many forms of disambiguation
in the properties. There are a few metadata values that are of particular
interest, specifically (in no particular order) title, type, units and verification.
Of particular note is the verification metadatum, this exposes a contextually
relevant set of certification criteria that lends validity to the property of
interest. Each property has a potentially unique set of metadata and
should be examined prior to using the property in your research. The following
are examples of using the schema directive: 

\url{API-Server?schema()}

or

\url{API-Server?schema(<keyword>)}

{\bf Nota Bene: Directives are not logical}

Directives do not support inter-directive logical relations. Intra-Directive
properties are not required to be logically related.  Any attempt to depend on
a logical relationship involving Directive keywords may not have the desired
result. Directive keywords are not interpreted in blocking context. Any attempt
to modify precedence by the use of blocking parentheses will likely cause a
failure in your request. Directive keywords may return an error state when used
as the sole operators in your request.

\begin{figure*}[t!]
\includegraphics[width=0.99\textwidth]{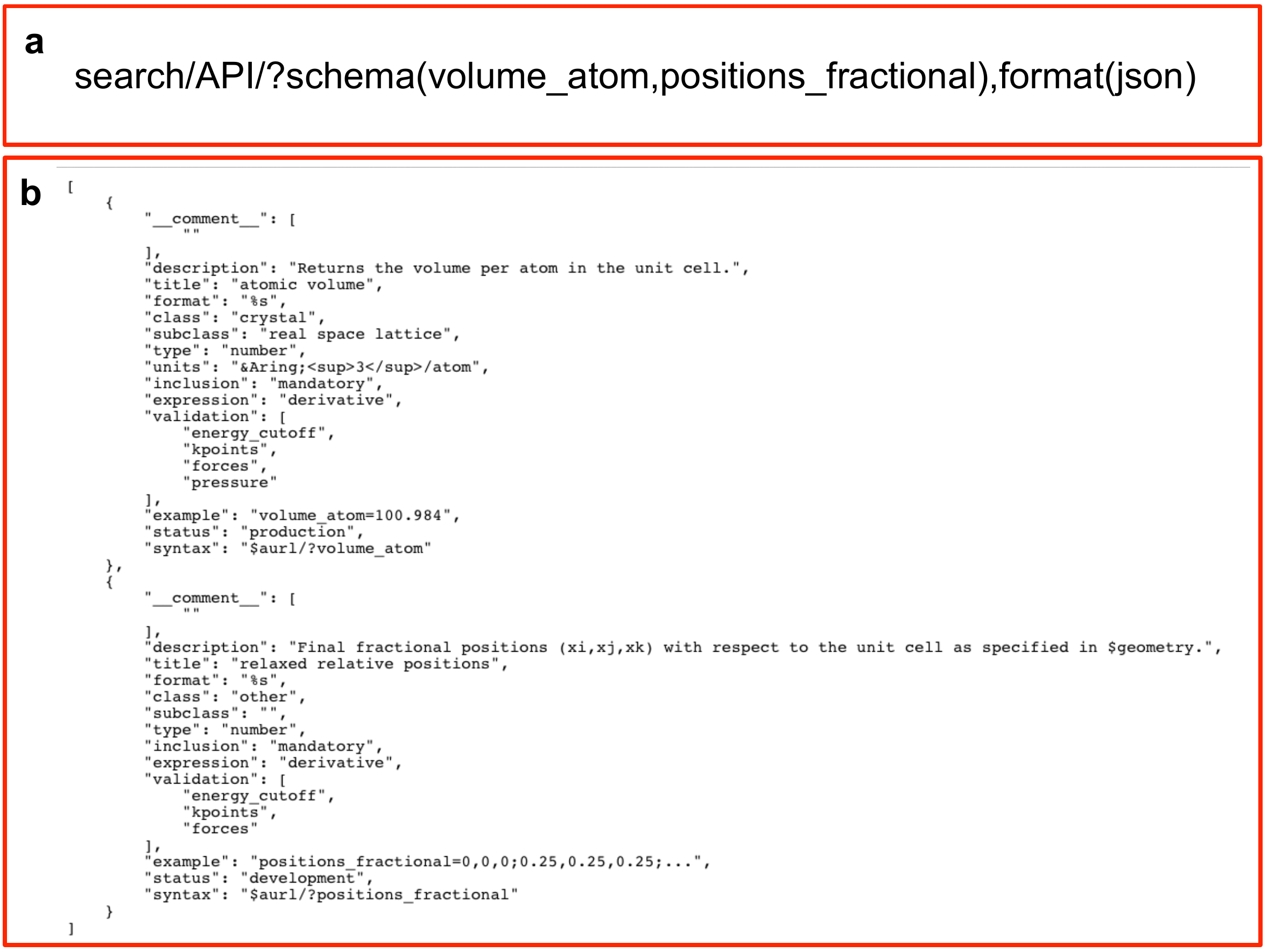}
\vspace{-4mm}
\caption{
\small 
{\bf (a)} The \AFLUX\ schema for a particular set of keywords can be retrieved using the ``schema'' Directive with the appropriate keywords as arguments. 
{\bf (b)} The schema includes information on the property type referenced by each keyword, the units that the property is provided in, and a
list of other keywords which can be used to validate the convergence and quality of the calculation used to generate that particular property.
}
\label{fig:AFLUX_schema}
\end{figure*}

\subsection{Implicit Procedures}

The first match keyword in the matchbook is used as the ordering
criteria for the response data set, so that the returned entries are
listed in increasing order of the property accessed by that keyword.

In the \AFLUX\ implementation the response always includes the compound
formula, the \AUID\ and the \AURL; i.e.  the keywords \verb|compound|, \verb|auid|
and \verb|aurl| are implicitly included in the query \cite{curtarolo:art92}. If any of these
properties are undesirable then they may be suppressed by application of the
suppression operator: e.g. the following will prevent the response from
producing the \verb|compound| property

\url{API-Server?<Matchbook>,$compound}.

The first data response is paged with 64 entries per page and retrieval of
subsequent pages require an explicit use of the paging directive.

\subsection{Matching Criteria}

The \AFLUX\ Search-\API\ can filter for search results that match a range of
different criteria, including exact matches and value ranges for both strings
and numerical data.

{\bf Numerical}

For scalar filtering of a single element we have three basic operations
and they are

\textless{} ``equality'' \textbar{} ``less'' \textbar{} ``more''
\textgreater{}

and their inverses, i.e.

\textless{} ``not equals'' \textbar{} ``not less'' \textbar{} ``not
more''\textgreater{}


The results on a numerical value associated with a particular a keyword
can be limited using the syntax shown in Table \ref{tab:numerical_filter}.
Note that the operations ``less than'' and ``greater than'' are
constructed by negating the ``greater than or equal to'' and ``less
than or equal to'' operations, respectively.







{\bf Strings}

For string matching we use a similar construction, as shown in Table \ref{tab:string_filter}.









Note that the string operator \textless{}`` ' ''\textgreater{} is optional
unless the matching string contains a reserved character.


{\bf Multiple Criterion}

The search described so far only demonstrates a single matching
condition, hereinafter referred to as a match. We also support the logical
construction of multiple matches, referred to as matchbooks. This is
done by the juxtaposition of matches with one of the two list separators,
namely \textless{}``:''\textgreater{} the logical \_or\_ operator and
\textless{}``,''\textgreater{} the logical \_and\_ operator. Precedence is
maintained by using nested lists. For instance, if we want to match two
criteria simultaneously the Matchbook might look as shown in the 
first row of Table \ref{tab:multiple_filter}. When matching multiple criteria,
 two forms are possible, as shown in the remaining rows of Table \ref{tab:multiple_filter}.








\section{Examples}
\label{examples}

\subsection{Search for electrically insulating heat sink material}
\label{example_phase_diagrams_2}

In this example we introduce the steps to screen for an electrically insulating heat sink material for use in nanoelectronics. We search for 
a material with a band gap in excess of 6eV which also has a high value of the lattice thermal conductivity. Therefore, as shown in 
Figure \ref{fig:search_format_example}(b), the Matchbook is constructed by querying two materials properties keywords. The first keyword
is \verb|agl_thermal_conductivity_300K|, which is the lattice thermal conductivity at 300K as calculated using the \AGL\ quasiharmonic
Debye-Gr{\"u}neisen model as implemented within the \AFLOW\ framework
\cite{curtarolo:art96, curtarolo:art115, Blanco_CPC_GIBBS_2004}. 
The second keyword is \verb|Egap|, the electronic structure band gap obtained by taking the difference between the conduction band
minimum and the valence band maximum, using the band structure calculated along the high-symmetry paths in reciprocal space
as defined by the \AFLOW\ Standard \cite{curtarolo:art58, curtarolo:art104}. In this case, we restrict the search to return only materials
with a band gap calculated to be greater than 6eV by searching for \verb|Egap(6*)|. The Directives come after the Matchbook. In this 
case there are two Directives, both related to the formatting of the returned output. The \verb|paging(-1,20)| Directive instructs the \API\ 
to return the first 20 results in descending order, while the \verb|format(html)| Directive instructs the \API\ to return the results in {\small HTML} format. 

The returned results for this search are shown in Figure \ref{fig:search_format_example}(c). In this case, 911 entries are found in
the \AFLOW\ data respository which match the requested search critreria. The entries are sorted in descending order of the value 
corresponding to the first property keyword in the matchbook, in this case \verb|agl_thermal_conductivity_300K|, due to 
the use of a negative value for the page number in the argument to the ``paging'' Directive. Therefore, the materials with the highest thermal conductivity 
which would be most suitable for this application appear at the top of the list, namely two different structural phases of the material BeO.

\section{Conclusion}
\label{conclusions}

In this article, we have presented \AFLUX, a \URI-based search \API\ extending the original materials data oriented
\AFLOW-\API. The syntax of the new materials language \AFLUX\ (\LUX\ for \AFLOW) enables the construction of
 complex search queries and facilitates the creation of remote operations on the \AFLOW.org repositories. 
The semantic of \AFLUX\ is transparent and easy to adapt to other materials genome initiatives, such as the Materials Project, NoMaD, {\small OQMD}, and AiiDA

\section{Acknowledgments}
\label{acknowledgements}
The authors thank Drs. O. Levy, I. Takeuchi, G. Hart, J. Carrete, J. Plata and N. Mingo for helpful discussions.
This work is partially supported by DOD-ONR (N00014-13-1-0635, N00014-11-1-0136, N00014-09-1-0921), 
NIST \#70NANB12H163 and by the Duke University---Center for Materials Genomics.
C.T. and S.C. acknowledge partial support by DOE (DE-AC02-05CH11231), specifically the BES program under Grant \#EDCBEE.
The consortium \AFLOW.org acknowledges the CRAY corporation for computational assistance.

\pagebreak

\appendix

\begin{widetext}

\section{Syntax Diagrams}

In the instantiation of an \AFLUX\ process we accept summons that conform to
the following syntax. The \LUX\ syntax is described in \ABNF\
(https://tools.ietf.org/html/rfc5234 and https://tools.ietf.org/html/rfc7405). 

\noindent
\verb|Summons =|
\begin{quote}
\small \verb|(Matchbook/Directive) *(Binal Directive)|
\end{quote}
\verb|Matchbook =|
\begin{quote}
\small \verb|[Unary-Not]([Unary-Mute]Datum-string"("Match")")/"("Matchbook")"/(Matchbook Binal Matchbook)|
\end{quote}
\verb|Match =|
\begin{quote}
\small \verb|[Unary-Not]([Unary-Loose](Datum-string/Datum-number)[Unary-Loose]/"("Match")")/(Match Binal Match)|
\end{quote}
\verb|Directive =|
\begin{quote}
\small \verb|[Unary-Mute]Datum-string"("(Datum-string/Datum-number) *(Binal(Datum-string/Datum-number))")"|
\end{quote}
\begin{figure}[h!]
\includegraphics[width=0.99\textwidth]{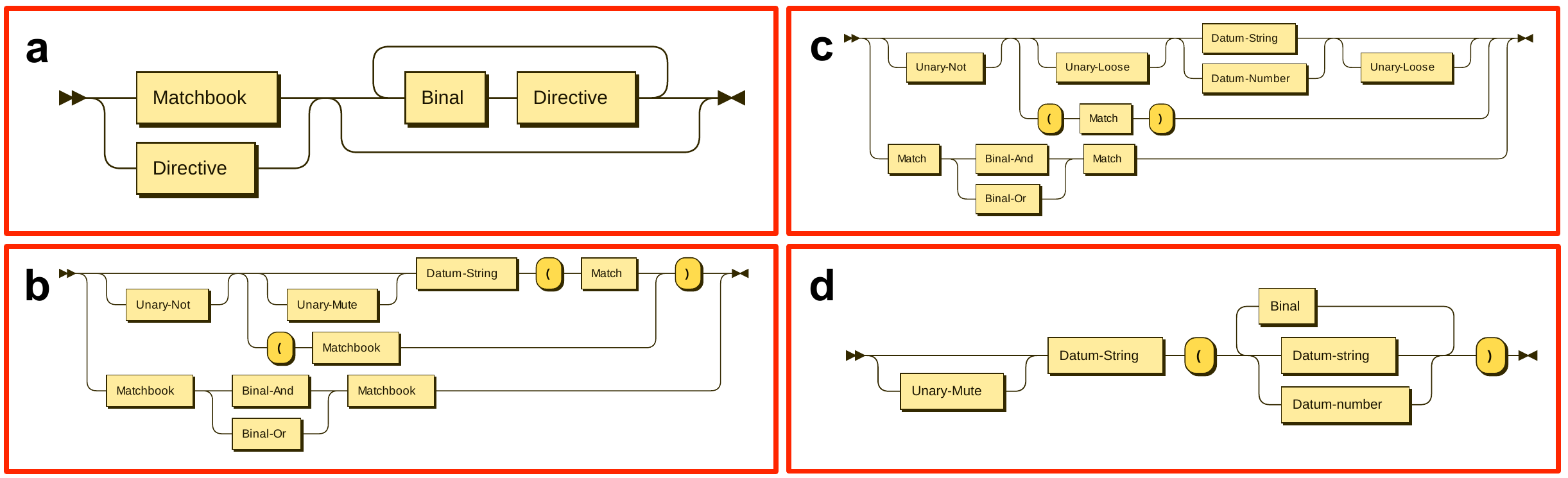}
\vspace{-4mm}
\caption{
\small Syntax diagrams for the \LUX\ search-API: (a) Summons, (b) Matchbook, (c) Match, (d) Directive.  
}
\label{fig:syntax_diagrams}
\end{figure}

\section{Table of Properties and \API\ Keywords}
\label{table_properties}
This section includes the new keywords added to the \AFLOW\ data \API\ since the release of version 1.0 as described 
in Ref. \onlinecite{curtarolo:art92}.
These include keywords which were added to facilitate verification 
as well as keywords that were
added to provide access to data calculated using the \AEL-\AGL\ methodology for thermomechanical properties \citeAFLOWAGL.
The \verb|kpoints| keyword has also been upgraded from version 1.0 to include the number of kpoints per line segment for the electronic
band structure calculation, and the new format is also described below. 
For each keyword, we list the description, type, inclusion policy and the retrieval syntax.

\def\description{\item {{\it Description.}\ }}
\def\type{\item {{\it Type.}\ }}
\def\example{\item {{\it Example.}\ }}
\def\inclusionmandatory{\item {{\it Inclusion.} \url{mandatory}}}
\def\inclusionoptional{\item {{\it Inclusion.} \url{optional}}}
\def\units{\item {{\it Units.}\ }}
\def\tol{\item {{\it Tolerance.}\ }}
\def\syntax{\item {{\it Request syntax.}\ }}

\def\ENERGYunit{e.g., eV or Ry if the calculations were performed with {\small VASP} \cite{vasp} or {\small QE} \cite{qe}, respectively}
\def\ENERGYunitatom{e.g., eV or Ry (eV/atom or Ry/atom) if the calculations were performed with {\small VASP} \cite{vasp} or {\small QE} \cite{qe}, respectively}
\def\LLLunits{e.g., \AA$^3$ or Bohr$^3$ (\AA$^3$/atom or Bohr$^3$/atom) if the calculations were performed with {\small VASP} \cite{vasp} or {\small QE} \cite{qe}, respectively}
\def\FORCEunit{e.g., eV/\AA\ or a.u if the calculations were performed with {\small VASP} \cite{vasp} or {\small QE} \cite{qe}, respectively}
\def\Lunits{e.g., \AA\ or a.u. (Bohr) if the calculations were performed with {\small VASP} \cite{vasp} or {\small QE} \cite{qe}, respectively}
\def\Punits{e.g., kbar or a.u. (Ry/Bohr) if the calculations were performed with {\small VASP} \cite{vasp} or {\small QE} \cite{qe}, respectively}
\def\Sunits{e.g., $\mu_B$ (Bohr magneton)}



\subsection{Optional materials keywords (alphabetic order)}
\label{optional_keywords_materials}

\begin{itemize}

\item
\verb|ael_bulk_modulus_reuss| 
\begin{itemize}
\description Returns AEL bulk modulus as calculated using the Reuss average.
\type \verb|number|.
\units GPa.
\example \verb|ael_bulk_modulus_reuss=105.315|
\syntax \verb|$aurl/?ael_bulk_modulus_reuss|
\end{itemize}

\item
\verb|ael_bulk_modulus_voigt| 
\begin{itemize}
\description Returns AEL bulk modulus as calculated using the Voigt average.
\type \verb|number|.
\units GPa.
\example \verb|ael_bulk_modulus_voigt=105.315|
\syntax \verb|$aurl/?ael_bulk_modulus_voigt|
\end{itemize}

\item
\verb|ael_bulk_modulus_vrh| 
\begin{itemize}
\description Returns AEL bulk modulus as calculated using the
Voigt-Reuss-Hill (VRH) average.
\type \verb|number|.
\units GPa.
\example \verb|ael_bulk_modulus_vrh=105.315|
\syntax \verb|$aurl/?ael_bulk_modulus_vrh|
\end{itemize}

\item
\verb|ael_elastic_anistropy| 
\begin{itemize}
\description Returns AEL elastic anisotropy.
\type \verb|number|.
\units dimensionless.
\example \verb|ael_elastic_anistropy=0.000816153|
\syntax \verb|$aurl/?ael_elastic_anisotropy|
\end{itemize}

\item
\verb|ael_poisson_ratio| 
\begin{itemize}
\description Returns AEL Poisson ratio.
\type \verb|number|.
\units dimensionless.
\example \verb|ael_poisson_ratio=0.21599|
\syntax \verb|$aurl/?ael_poisson_ratio|
\end{itemize}

\item
\verb|ael_shear_modulus_reuss| 
\begin{itemize}
\description Returns AEL shear modulus as calculated using the Reuss average.
\type \verb|number|.
\units GPa.
\example \verb|ael_shear_modulus_reuss=73.7868|
\syntax \verb|$aurl/?ael_shear_modulus_reuss|
\end{itemize}

\item
\verb|ael_shear_modulus_voigt| 
\begin{itemize}
\description Returns AEL shear modulus as calculated using the Voigt average.
\type \verb|number|.
\units GPa.
\example \verb|ael_shear_modulus_voigt=73.7989|
\syntax \verb|$aurl/?ael_shear_modulus_voigt|
\end{itemize}

\item
\verb|ael_shear_modulus_vrh| 
\begin{itemize}
\description Returns AEL shear modulus as calculated using the
Voigt-Reuss-Hill (VRH) average.
\type \verb|number|.
\units GPa.
\example \verb|ael_shear_modulus_vrh=73.7929|
\syntax \verb|$aurl/?ael_shear_modulus_vrh|
\end{itemize}

\item
\verb|ael_speed_of_sound_average| 
\begin{itemize}
\description Returns AEL average speed of sound calculated from the transverse and longitudinal speeds of sound.
\type \verb|number|.
\units m/s.
\example \verb|ael_speed_of_sound_average=500.0|
\syntax \verb|$aurl/?ael_speed_of_sound_average|
\end{itemize}

\item
\verb|ael_speed_of_sound_longitudinal| 
\begin{itemize}
\description Returns AEL speed of sound in the longitudinal direction.
\type \verb|number|.
\units m/s.
\example \verb|ael_speed_of_sound_longitudinal=500.0|
\syntax \verb|$aurl/?ael_speed_of_sound_longitudinal|
\end{itemize}

\item
\verb|ael_speed_of_sound_transverse| 
\begin{itemize}
\description Returns AEL speed of sound in the transverse direction.
\type \verb|number|.
\units m/s.
\example \verb|ael_speed_of_sound_transverse=500.0|
\syntax \verb|$aurl/?ael_speed_of_sound_transverse|
\end{itemize}

\item
\verb|agl_acoustic_debye| 
\begin{itemize}
\description Returns AGL acoustic Debye temperature.
\type \verb|number|.
\units K.
\example \verb|agl_acoustic_debye=492|
\syntax \verb|$aurl/?agl_acoustic_debye|
\end{itemize}

\item
\verb|agl_bulk_modulus_isothermal_300K| 
\begin{itemize}
\description Returns AGL isothermal bulk modulus at 300K and zero pressure.
\type \verb|number|.
\units GPa.
\example \verb|agl_bulk_modulus_isothermal_300K=96.6|
\syntax \verb|$aurl/?agl_bulk_modulus_isothermal_300K|
\end{itemize}

\item
\verb|agl_bulk_modulus_static_300K| 
\begin{itemize}
\description Returns AGL static bulk modulus at 300K and zero pressure.
\type \verb|number|.
\units GPa.
\example \verb|agl_bulk_modulus_static_300K=99.59|
\syntax \verb|$aurl/?agl_bulk_modulus_static_300K|
\end{itemize}

\item
\verb|agl_debye| 
\begin{itemize}
\description Returns AGL Debye temperature.
\type \verb|number|.
\units K.
\example \verb|agl_debye=620|
\syntax \verb|$aurl/?agl_debye|
\end{itemize}

\item
\verb|agl_gruneisen| 
\begin{itemize}
\description Returns AGL Gr{\"u}neisen parameter.
\type \verb|number|.
\units dimensionless.
\example \verb|agl_gruneisen=2.06|
\syntax \verb|$aurl/?agl_gruneisen|
\end{itemize}

\item
\verb|agl_heat_capacity_Cv_300K| 
\begin{itemize}
\description Returns AGL heat capacity at constant volume (C$_V$) at 300K and zero pressure.
\type \verb|number|.
\units k$_\mathrm{B}$/cell.
\example \verb|agl_heat_capacity_Cv_300K=4.901|
\syntax \verb|$aurl/?agl_heat_capacity_Cv_300K|
\end{itemize}

\item
\verb|agl_heat_capacity_Cp_300K| 
\begin{itemize}
\description Returns AGL heat capacity at constant pressure (C$_p$) at 300K and zero pressure.
\type \verb|number|.
\units k$_\mathrm{B}$/cell.
\example \verb|agl_heat_capacity_Cp_300K=5.502|
\syntax \verb|$aurl/?agl_heat_capacity_Cp_300K|
\end{itemize}

\item
\verb|agl_poisson_ratio_source| 
\begin{itemize}
\description Returns source of Poisson ratio used to calculate Debye temperature in AGL. Possible sources include \verb|ael|, in
which case the Poisson ratio was calculated from first principles using AEL; \verb|empirical|, in which case the value was taken 
from the literature; and \verb|Cauchy_ratio_0.25|, in which case the default value of 0.25 of the Poisson ratio of a Cauchy solid
was used.
\type \verb|string|.
\example \verb|agl_poisson_ratio_source=ael|
\syntax \verb|$aurl/?agl_poisson_ratio_source|
\end{itemize}

\item
\verb|agl_thermal_conductivity_300K| 
\begin{itemize}
\description Returns AGL thermal conductivity at 300K.
\type \verb|number|.
\units W/m*K.
\example \verb|agl_thermal_conductivity_300K=24.41|
\syntax \verb|$aurl/?agl_thermal_conductivity_300K|
\end{itemize}

\item
\verb|agl_thermal_expansion_300K| 
\begin{itemize}
\description Returns AGL thermal expansion at 300K and zero pressure.
\type \verb|number|.
\units 1/K.
\example \verb|agl_thermal_expansion_300K=4.997e-05|
\syntax \verb|$aurl/?agl_thermal_expansion_300K|
\end{itemize}

\item
\verb|delta_electronic_energy_final|
\begin{itemize}
\description Convergence energy threshold for the electronic SCF loop.
\type \verb|number|.
\example \verb|delta_electronic_energy_final=0.000071416|
\syntax \verb|$aurl/?delta_electronic_energy_final|
\end{itemize}

\item
\verb|delta_electronic_energy_threshold|
\begin{itemize}
\description Convergence energy threshold for the electronic SCF loop.
\type \verb|number|.
\example \verb|delta_electronic_energy_threshold=0.0001|
\syntax \verb|$aurl/?delta_electronic_energy_threshold|
\end{itemize}

\item
\verb|kpoints|
\begin{itemize}
\description Set of {\bf k}-point meshes uniquely identifying the various steps of the calculations,
e.g.\ relaxation, static and electronic band structure (specifying the {\bf k}-space symmetry points of
the structure and the number of points per path segment).
\type Set of \verb|numbers| and \verb|strings| separated by ``,'' and ``;''.
\example \verb|kpoints=10,10,10;16,16,16;\Gamma-X,X-W,W-K,K-\Gamma,\Gamma-L,L-U,U-W,W-L,L-K,U-X;20|
\syntax \verb|$aurl/?kpoints|
\end{itemize}

\item
\verb|nkpoints|
\begin{itemize}
\description Number of k-points used for the calculation.
\type \verb|number|.
\example \verb|nkpoints=2197|
\syntax \verb|$aurl/?nkpoints|
\end{itemize}

\item
\verb|nkpoints_irreducible|
\begin{itemize}
\description Number of k-points in the irreducible Brillouin zone.
\type \verb|number|.
\example \verb|nkpoints=2197|
\syntax \verb|$aurl/?nkpoints|
\end{itemize}

\item
\verb|pressure_residual|
\begin{itemize}
\description Returns the residual pressure for the simulation, i.e. the difference between the pressure specified in the input and the actual pressure achieved for the relaxation.
\type \verb|number|.
\units Natural units of the \verb|$code|, \Punits.
\example \verb|pressure_residual=10.0|
\syntax \verb|$aurl/?pressure_residual|
\end{itemize}

\item
\verb|Pulay_stress|
\begin{itemize}
\description Returns the Pulay stress for the simulation.
\type \verb|number|.
\units Natural units of the \verb|$code|, \Punits.
\example \verb|Pulay_stress=10.0|
\syntax \verb|$aurl/?Pulay_stress|
\end{itemize}

\item
\verb|stress_tensor|
\begin{itemize}
\description Returns the stress tensor obtained for the simulation.
\type List of 9 \verb|numbers| separated by commas, giving the elements of the stress tensor in the form $S_{xx}, S_{xy}, S_{xz}, S_{yx}, S_{yy}, S_{yz}, S_{zx}, S_{zy}, S_{zz}$.
\units Natural units of the \verb|$code|, \Punits.
\example \verb|stress_tensor=0.74,0,0,0,0.74,-0,0,-0,-4.42|
\syntax \verb|$aurl/?stress_tensor|
\end{itemize}

\end{itemize}

\end{widetext}



\section*{References}
\small
\begin{thebibliography}{10}

\bibitem{aflowPAPER}
S.~Curtarolo, W.~Setyawan, G.~L.~W. Hart, M.~Jahn\'{a}tek, R.~V. Chepulskii,
  R.~H. Taylor, S.~Wang, J.~Xue, K.~Yang, O.~Levy, M.~J. Mehl, H.~T. Stokes,
  D.~O. Demchenko, and D.~Morgan, \emph{{AFLOW}: An automatic framework for
  high-throughput materials discovery}, Comput.\ Mater.\ Sci. \textbf{58},
  218--226 (2012).

\bibitem{curtarolo:art110}
K.~Yang, C.~Oses, and S.~Curtarolo, \emph{Modeling Off-Stoichiometry Materials
  with a High-Throughput {\it Ab-Initio} Approach}, Chem.\ Mater.  (2016).

\bibitem{curtarolo:art104}
C.~E. Calderon, J.~J. Plata, C.~Toher, C.~Oses, O.~Levy, M.~Fornari, A.~Naturen,
  M.~J. Mehl, G.~L.~W. Hart, M.~{Buongiorno~Nardelli}, and S.~Curtarolo,
  \emph{The {AFLOW} standard for high-throughput materials science
  calculations}, Comput.\ Mater.\ Sci. \textbf{108 Part A}, 233--238 (2015).

\bibitem{curtarolo:art63}
O.~Levy, M.~Jahn\'{a}tek, R.~V. Chepulskii, G.~L.~W. Hart, and S.~Curtarolo,
  \emph{Ordered Structures in {R}henium Binary Alloys from First-Principles
  Calculations}, J.\ Am.\ Chem.\ Soc. \textbf{133}, 158--163 (2011).

\bibitem{curtarolo:art57}
O.~Levy, G.~L.~W. Hart, and S.~Curtarolo, \emph{Structure maps for hcp metals
  from first-principles calculations}, Phys.\ Rev.\ B \textbf{81}, 174106
  (2010).

\bibitem{curtarolo:art49}
O.~Levy, G.~L.~W. Hart, and S.~Curtarolo, \emph{Uncovering Compounds by Synergy
  of Cluster Expansion and High-Throughput Methods}, J.\ Am.\ Chem.\ Soc.
  \textbf{132}, 4830--4833 (2010).

\bibitem{monsterPGM}
G.~L.~W. Hart, S.~Curtarolo, T.~B. Massalski, and O.~Levy, \emph{Comprehensive
  Search for New Phases and Compounds in Binary Alloy Systems Based on
  {P}latinum-Group Metals, Using a Computational First-Principles Approach},
  Phys.\ Rev.\ X \textbf{3}, 041035 (2013).

\bibitem{curtarolo:art84}
J.~Carrete, W.~Li, N.~Mingo, S.~Wang, and S.~Curtarolo, \emph{Finding
  Unprecedentedly Low-Thermal-Conductivity Half-Heusler Semiconductors via
  High-Throughput Materials Modeling}, Phys.\ Rev.\ X \textbf{4}, 011019
  (2014).

\bibitem{curtarolo:art85}
J.~Carrete, N.~Mingo, S.~Wang, and S.~Curtarolo, \emph{Nanograined
  Half-{H}eusler Semiconductors as Advanced Thermoelectrics: An Ab Initio
  High-Throughput Statistical Study}, Adv.\ Func.\ Mater. \textbf{24},
  7427--7432 (2014).

\bibitem{curtarolo:art94}
O.~Isayev, D.~Fourches, E.~N. Muratov, C.~Oses, K.~Rasch, A.~Tropsha, and
  S.~Curtarolo, \emph{Materials Cartography: Representing and Mining Materials
  Space Using Structural and Electronic Fingerprints}, Chem.\ Mater.
  \textbf{27}, 735--743 (2015).

\bibitem{curtarolo:art120}
A.~van Roekeghem, J.~Carrete, C.~Oses, S.~Curtarolo, and N.~Mingo, \emph{High
  throughput thermal conductivity of high temperature solid phases: The case of
  oxide and fluoride perovskites}, Phys.\ Rev.\ X  (2016).

\bibitem{curtarolo:art81}
S.~Curtarolo, G.~L.~W. Hart, M.~{Buongiorno~Nardelli}, N.~Mingo, S.~Sanvito,
  and O.~Levy, \emph{The high-throughput highway to computational materials
  design}, Nature\ Mater. \textbf{12}, 191--201 (2013).

\bibitem{aflowlibPAPER}
S.~Curtarolo, W.~Setyawan, S.~Wang, J.~Xue, K.~Yang, R.~H. Taylor, L.~J.
  Nelson, G.~L.~W. Hart, S.~Sanvito, M.~{Buongiorno~Nardelli}, N.~Mingo, and
  O.~Levy, \emph{{AFLOWLIB.ORG}: A distributed materials properties repository
  from high-throughput {\it ab initio} calculations}, Comput.\ Mater.\ Sci.
  \textbf{58}, 227--235 (2012).

\bibitem{curtarolo:art58}
W.~Setyawan and S.~Curtarolo, \emph{High-throughput electronic band structure
  calculations: Challenges and tools}, Comput.\ Mater.\ Sci. \textbf{49},
  299--312 (2010).

\bibitem{curtarolo:art92}
R.~H. Taylor, F.~Rose, C.~Toher, O.~Levy, K.~Yang, M.~{Buongiorno~Nardelli},
  and S.~Curtarolo, \emph{A {REST}ful {API} for exchanging Materials Data in
  the {AFLOWLIB}.org consortium}, Comput.\ Mater.\ Sci. \textbf{93}, 178--192
  (2014).

\bibitem{nomad}
M.~Scheffler, C.~Draxl, and {Computer Center of the Max-Planck Society,
  Garching}, \emph{The {NoMaD} Repository}, http://nomad-repository.eu (2014).

\bibitem{materialsproject.org}
A.~Jain, G.~Hautier, C.~J. Moore, S.~P. Ong, C.~C. Fischer, T.~Mueller, K.~A.
  Persson, and G.~Ceder, \emph{A high-throughput infrastructure for density
  functional theory calculations}, Comput.\ Mater.\ Sci. \textbf{50},
  2295--2310 (2011).

\bibitem{Saal_JOM_2013}
J.~E. Saal, S.~Kirklin, M.~Aykol, B.~Meredig, and C.~Wolverton, \emph{Materials
  Design and Discovery with High-Throughput Density Functional Theory: The
  {O}pen {Q}uantum {M}aterials {D}atabase ({OQMD})}, JOM \textbf{65},
  1501--1509 (2013).

\bibitem{aiida.net}
G.~Pizzi, A.~Cepellotti, R.~Sabatini, N.~Marzari, and B.~Kozinsky, {\it AiiDA},
  {\sf http://www.aiida.net}  (2016).

\bibitem{Pizzi_AiiDA_2016}
G.~Pizzi, A.~Cepellotti, R.~Sabatini, N.~Marzari, and B.~Kozinsky, \emph{AiiDA:
  automated interactive infrastructure and database for computational science},
  Comp.\ Mat.\ Sci. \textbf{111}, 218--230 (2016).

\bibitem{RFC3986}
T.~Berners-Lee, R.~Fielding, and L.~Masinter, \emph{Uniform Resource Identifier
  (URI): Generic Syntax}  (2005).

\bibitem{RFC5234}
E.~B. D.~Crocker and P.~Overell, \emph{Augmented BNF for Syntax Specifications:
  ABNF}  (2008).

\bibitem{curtarolo:art96}
C.~Toher, J.~J. Plata, O.~Levy, M.~{de~Jong}, M.~D. Asta,
  M.~{Buongiorno~Nardelli}, and S.~Curtarolo, \emph{High-throughput
  computational screening of thermal conductivity, {D}ebye temperature, and
  {G}r{\"u}neisen parameter using a quasiharmonic {D}ebye Model}, Phys.\ Rev.\
  B \textbf{90}, 174107 (2014).

\bibitem{curtarolo:art115}
C.~Toher, C.~Oses, J.~J. Plata, D.~Hicks, F.~Rose, O.~Levy, M.~{de Jong}, M.~D.
  Asta, M.~Fornari, M.~{Buongiorno Nardelli}, and S.~Curtarolo, \emph{Combining
  the AFLOW GIBBS and Elastic Libraries for efficiently and robustly screening
  thermo-mechanical properties of solids}, arXiv:1611.05714  (2016).

\bibitem{Blanco_CPC_GIBBS_2004}
M.~A. Blanco, E.~Francisco, and V.~Lua{\~n}a, \emph{GIBBS: isothermal-isobaric
  thermodynamics of solids from energy curves using a quasi-harmonic Debye
  model}, Comput.\ Phys.\ Commun. \textbf{158}, 57--72 (2004).

\bibitem{vasp}
G.~Kresse and J.~Furthm\"uller, \emph{Efficient iterative schemes for {\it ab
  initio} total-energy calculations using a plane-wave basis set}, Phys.\ Rev.\
  B \textbf{54}, 11169--11186 (1996).

\bibitem{qe}
P.~Giannozzi, S.~Baroni, N.~Bonini, M.~Calandra, R.~Car, C.~Cavazzoni,
  D.~Ceresoli, G.~L. Chiarotti, M.~Cococcioni, I.~Dabo, A.~{Dal Corso}, S.~{de
  Gironcoli}, S.~Fabris, G.~Fratesi, R.~Gebauer, U.~Gerstmann, C.~Gougoussis,
  A.~Kokalj, M.~Lazzeri, L.~Martin-Samos, N.~Marzari, F.~Mauri, R.~Mazzarello,
  S.~Paolini, A.~Pasquarello, L.~Paulatto, C.~Sbraccia, S.~Scandolo,
  G.~Sclauzero, A.~P. Seitsonen, A.~Smogunov, P.~Umari, and R.~M. Wentzcovitch,
  \emph{QUANTUM ESPRESSO: a modular and open-source software project for
  quantum simulations of materials}, J.\ Phys.:\ Condens.\ Matter \textbf{21},
  395502 (2009).

\end{thebibliography}
\end{document}